# Robust zero resistance in a superconducting high entropy alloy against pressure up to 190 GPa

Jing Guo[1]*, Honghong Wang[1,4]*, Fabian von Rohr[2]*, Zhe Wang[1,4], Shu Cai[1,4], Yazhou Zhou[1,4], Ke Yang[3], Aiguo Li[3], Sheng Jiang[3], Qi Wu[1], Robert J Cava[2], Liling Sun[1,4,5]†

[1]Institute of Physics, Chinese Academy of Sciences, Beijing 100190, China
[2]Department of Chemistry, Princeton University, Princeton, New Jersey 08544, USA
[3]Shanghai Synchrotron Radiation Facilities, Shanghai Institute of Applied Physics, Chinese Academy of Sciences, Shanghai 201204, China
[4]University of Chinese Academy of Sciences, Beijing 100190, China
[5]Collaborative Innovation Center of Quantum Matter, Beijing, 100190, China

We report the observation of extraordinarily robust zero-resistance superconductivity in the pressurized $(TaNb)_{0.67}(HfZrTi)_{0.33}$ high entropy alloy - a new kind of material with a body-centered cubic crystal structure made from five randomly distributed transition metal elements. The transition to superconductivity ($T_C$) increases from an initial temperature of 7.7 K at ambient pressure to 10 K at ~ 60 GPa, and then slowly decreases to 9 K by 190.6 GPa, a pressure that falls within that of the outer core of the earth. We infer that the continuous existence of the zero-resistance superconductivity from one atmosphere up to such a high pressure requires a special combination of electronic and mechanical characteristics. This high entropy alloy superconductor thus may have a bright future for applications under extreme conditions, and also poses a challenge for understanding the underlying quantum physics.

High entropy alloys (HEAs) are a new class of materials that are composed of multiple transition metal elements in equimolar or near equimolar ratios[1,2]. The diverse elements in HEAs are arranged randomly on the crystallographic positions in a simple lattice, and thus have been referred to as a metallic glass on an ordered lattice. By applying this concept, many HEAs have been found in disordered solid solution phases with body-centered cubic, hexagonal closest-packed and face-centered cubic crystal structures[3-6].

In many respects, HEAs display novel properties, including ultrahigh fracture toughness at cryogenic temperatures[7,8], excellent specific strength[9], and superior mechanical performance at high temperatures[10]. In addition to their promising mechanical properties, some HEAs also exhibit interesting electronic properties - [TaNb]$_{1-x}$(ZrHfTi)$_x$ HEAs were found to display superconductivity for example[11,12]. The combination of the promising physical properties found in the HEAs points to great potential for application.

Pressure is one of the variables that can uncover unexpected phenomena and properties[13-16]. For superconductors in particular, the pressure-induced enhancement of critical transition temperatures in copper-oxide and iron-pnictide superconductors[17-20], the reemergence of superconductivity in the alkaline iron selenide[21] and heavy fermion superconductors[22], pressure-induced superconductivity in H$_3$S[23-25] and elements[26,27], are examples. Therefore, looking for new phenomena in the superconducting HEA under pressure is of great interest. Here we report the first high pressure studies on the superconducting HEA (TaNb)$_{0.67}$(HfZrTi)$_{0.33}$, which has a critical transition temperature to the superconducting state ($T_C$) of about 7.8 K at ambient pressure[11,12]. Our observations demonstrate that this alloy exhibits extraordinarily robust superconductivity - its zero resistance superconducting state is still achieved even at a pressure of 190.6 GPa, or almost 2 Megabars (1 Mbar =$10^{11}$ pascal), a pressure like that within the outer core of the earth. Such a superconductor with a highly robust zero resistance state, existing continuously from one atmosphere to geological pressures, is extremely unusual; and in fact is unique to the best of our knowledge. We attribute this surprising behavior to the stable crystal structure of the HEA combined with the apparent robustness of its electronic structure against very large amounts of lattice compression.

Figure 1a shows the temperature dependence of the electrical resistance at ambient pressure for a $(TaNb)_{0.67}(HfZrTi)_{0.33}$ sample. A sharp drop to a zero resistance superconducting state is observed at ~7.7 K (we define $T_C$ throughout as the temperature where the resistance changes from a finite value to zero), consistent with the results reported in Ref. 12. Applying a magnetic field on the sample shows that its superconducting transition temperature ($T_C$) systematically shifts to lower temperature (inset of Fig.1a), as expected. Temperature-dependent constant-current magnetic susceptibility characterization was also performed for the ambient-pressure sample. As shown in Fig.1b, a strong diamagnetic response is observed starting at 7.6 K, indicative of a bulk superconducting nature.

High-pressure resistance measurements were performed for four samples that were cut from the material used as the standard for the superconductivity at ambient pressure. The electrical resistance measurements for these samples were performed between 4 K and 300 K. Figure 2a shows the temperature dependence of electrical resistance measured in the pressure range of 1.02 - 58.2 GPa for one of the samples. It is seen that the superconducting transitions of the sample subjected to different pressures are sharp and the zero resistance state remains present throughout the full range of pressures applied. Zooming in on the resistance in the low temperature regime (Fig.2b), we find that the superconducting transition temperature $T_c$ shifts to higher temperature upon increasing pressure. Consistent results are obtained from another sample in the pressure range of 0.3 GPa – 30.1 GPa, as shown in Fig.3. To investigate the superconducting behavior at higher pressures, we carried out resistance measurements for the third sample over the wide pressure range of 1.5 - 179.2 GPa (Fig.2c-2d). As shown in Fig.2c, the superconducting state still survives at a pressure as high as 179.2 GPa. Surprisingly, the zero-resistance state still exists at this high pressure (Fig.2d). The magnetic field dependence of the superconducting transition measured at 103.7 GPa and 179.2 GPa respectively further confirm the superconducting nature (Fig.2e and Fig. 2f). Furthermore, our measurements up to 190.6 GPa on the fourth sample show the reproducibility of the results (Fig.3). To our knowledge, this is the first discovery of such robust zero-resistance superconductivity in a material from one atmosphere up to near 2 Mbar, a pressure found inside the outer core of the earth.

We summarize our experimental results, obtained from samples measured in the four independent runs, in the pressure-temperature phase diagram shown in Fig.3. It is seen that the one atmosphere superconducting $T_{Cs}$ of these samples are almost the same at approximately 7.7 K - different from the $T_C$ of any element included in the superconducting HEA investigated[28,29] and clearly shown in previous studies to be a bulk property of the HEA[11,12]. $T_C$ increases with pressure for the $(TaNb)_{0.67}(HfZrTi)_{0.33}$ HEA, exhibiting a slow increase from its ambient-pressure value of ~7.7 K to 10 K at ~ 60 GPa. On further increasing the pressure, $T_C$ remains almost unchanged until ~100 GPa and then shows a slight decline until the pressure of 190.6 GPa where the $T_C$ is about 9 K. Attempts to apply an even higher pressure than 190.6 GPa led to breaking of the diamonds in the pressure cell, unfortunately, so we do not know whether the zero resistance state survives to the pressure of earth's inner core.

High pressure synchrotron X-ray diffraction (XRD) measurements on the $(TaNb)_{0.67}(HfZrTi)_{0.33}$ HEA were performed at beamline 15U of the Shanghai Synchrotron Radiation Facility. The results of two independent experiments show that the superconducting HEA does not undergo a structural phase transition up to pressures of ~ 96 GPa where the volume is compressed by ~28% (Fig.4a-4b and Fig.S1 of Supplementary Information), but its body-centered cubic (BCC) structure is still maintained. Further, if the change of the volume is linearly extrapolated up to 190 GPa which is the highest investigated pressure of resistance measurements, the volume is compressed by ~ 53%.

An analysis of the data (Fig.4c) shows that the pressure dependence of the superconducting critical temperature $T_C$ of the body-centered-cubic superconducting

HEA $(TaNb)_{0.67}(HfZrTi)_{0.33}$ is quite distinct from those of the body-centered cubic superconductors Nb and Ta, which are the major elemental constituents of the HEA material (The superconducting $T_C$ of elemental Ta has been measured to higher pressures than are found in the literature as part of the current study (Fig.S2 of the SI). The $T_C$ of the HEA superconductor continuously increases up to 96 GPa, distinct from what is observed in both Nb and Ta in the course of their exposure to pressure[28,29]. Thus the superconducting HEA $(TaNb)_{0.67}(HfZrTi)_{0.33}$ is clearly distinct, and not just a diluted form of Nb or Ta.

To further evaluate the potential applicability of the HEA superconductor under extreme conditions, we also estimate the value of the upper critical magnetic field ($H_{c2}$, the magnetic field at which superconductivity disappears) for the pressurized material by using the Werthamer-Helfand-Hohenberg (WHH) formula[30]:

$$H_{c2}^{WHH}(0) = -0.693 T_C (dH_{C2}/dT)_{T=Tc}$$

Figure 4d shows plot of $H_{c2}$ versus $T_C$ obtained at different pressures, in which the dotted lines represent the WHH fits. The estimated values of the upper critical fields at zero temperature are ~8T at ambient pressure, ~ 4T at 100 GPa and ~ 2T at 179.2 GPa (inset of Fig.4d). The very robust zero-resistance state and robust upper critical fields from ambient pressure to pressures as high as that of the earth's outer core, together with the high compressibility (the volume is compressed by ~28% at 96 GPa, and the linear extrapolated change of the volume up to 190 GPa is ~ 53%), make the superconducting HEA a promising candidate for new applications and also pose a challenge for verifying the known superconductivity theory and developing new one.

**Acknowledgements**

The work in China was supported by the National Key Research and Development Program of China (Grant No. 2017YFA0302900, 2016YFA0300300 and 2017YFA0303103), the NSF of China (Grants No. 91321207, No. 11427805, No. U1532267, No. 11604376), the Strategic Priority Research Program (B) of the Chinese Academy of Sciences (Grant No. XDB07020300). The work at Princeton was supported by the Gordon and Betty Moore Foundation EPiQS initiative, grant GBMF-4412.

**Author information**

†Corresponding authors

llsun@iphy.ac.cn

* These authors contributed equally.

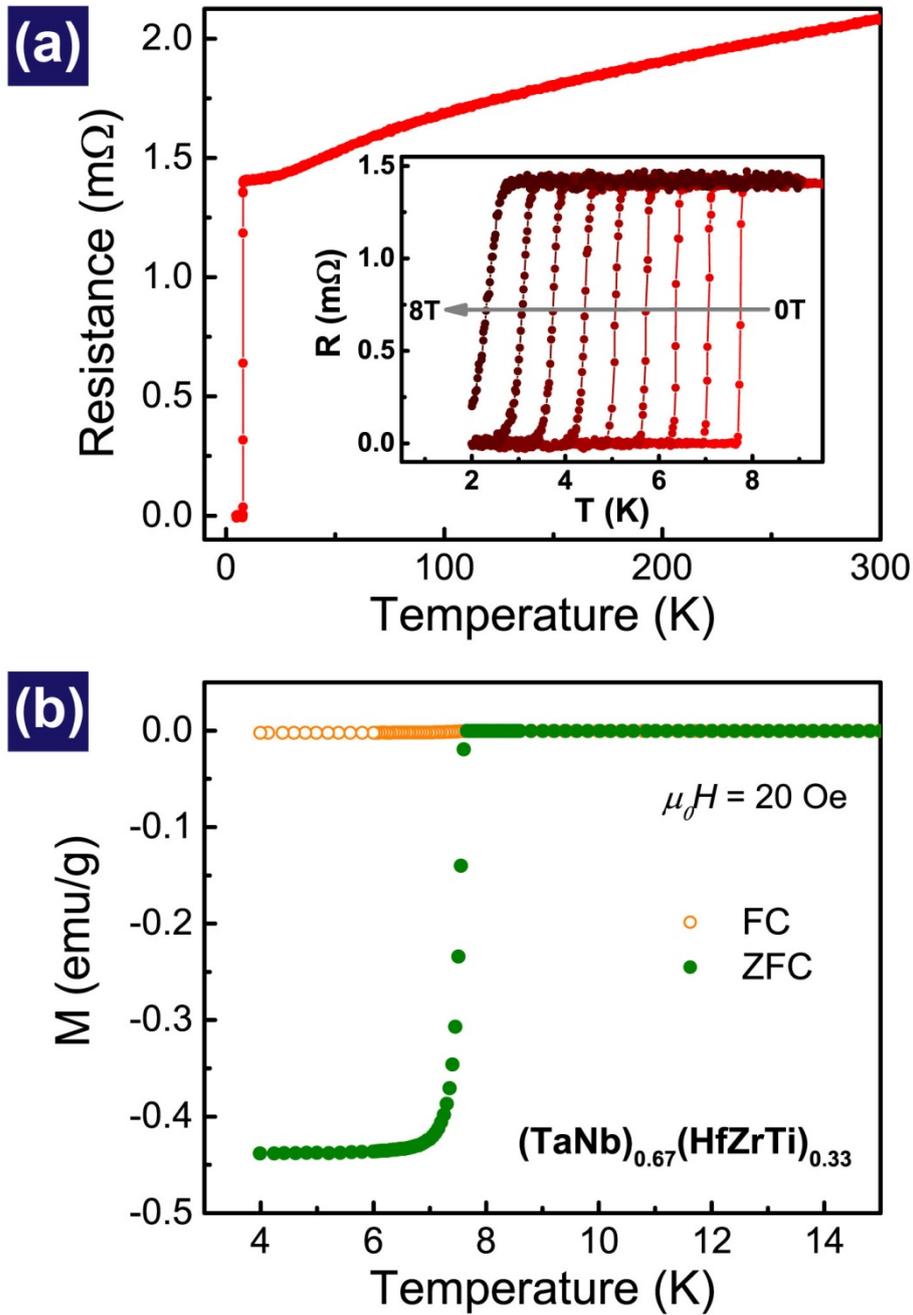

**Figure 1 Ambient-pressure properties of the superconducting high entropy alloy (TaNb)$_{0.67}$(HfZrTi)$_{0.33}$.** (a) Resistance measured in the temperature range from 2 K to 300 K. The inset shows the magnetic field dependence of the superconducting transition in fields from 0T to 8T. (b) Zero-field cooling (ZFC) and field-cooling (FC)

magnetization of the HEA in the vicinity of superconducting transition – the difference between the FC and ZFC curves is evidence for significant flux pinning on the material.

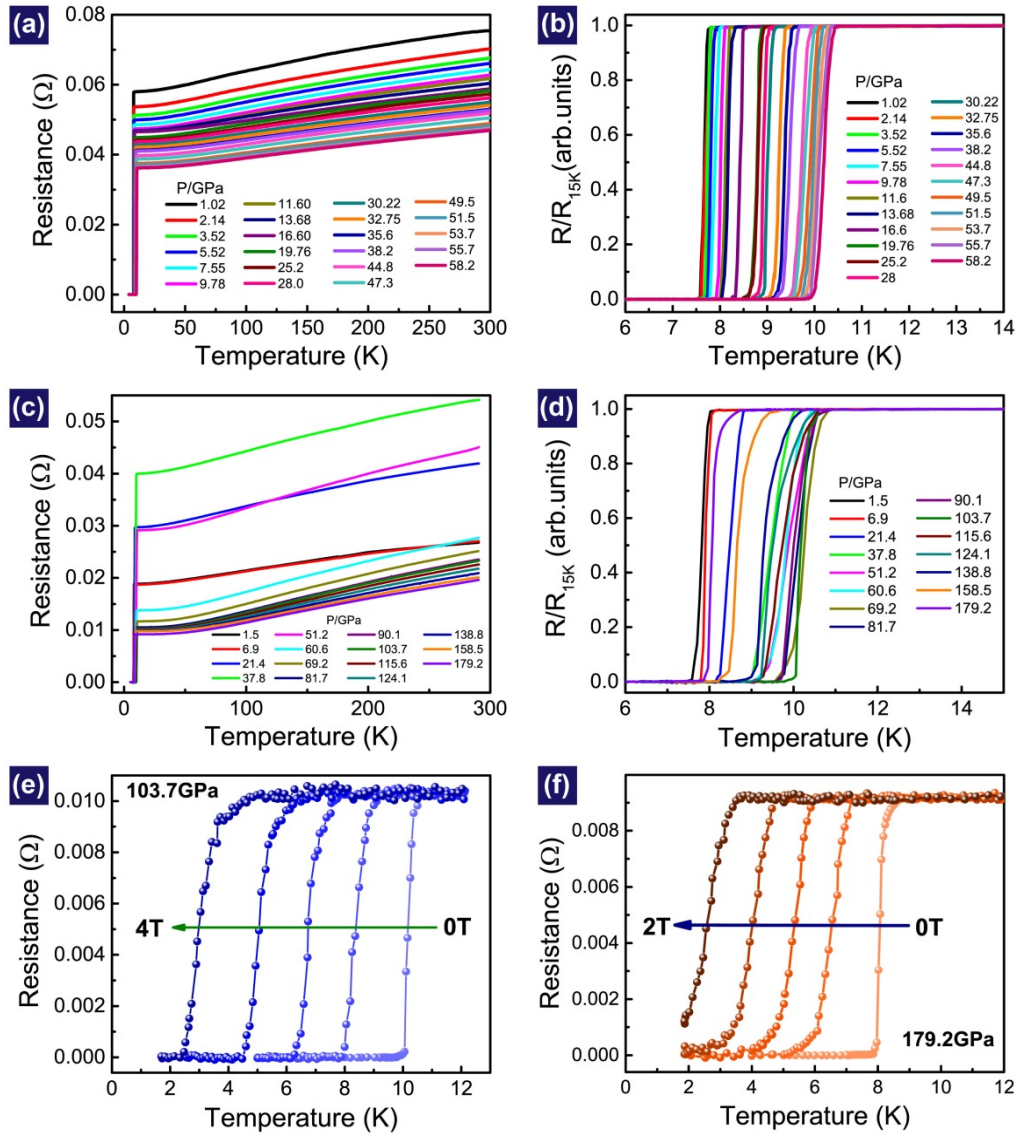

**Figure 2 Characterization of the superconducting transition of the HEA at high pressures**. (a) Resistance versus temperature obtained from the measurements in the pressure range of 1.02-58.2 GPa, over a wide temperature range. (b) Detail of the normalized resistance at low temperatures for the data in figure 2a, clearly showing

the effect of pressure on the resistance through the superconducting transition and the maintenance of the zero resistance state over a very wide pressure range. (c) Temperature dependence of resistance measured on another sample in the pressure range of 1.5-179.2 GPa over a wide pressure range. (d) Enlarged view of normalized resistance in figure 2c, illustrating the superconducting transition at different pressures and in particular the robust zero-resistance state up to the pressure of 179.2 GPa. (e) and (f) Magnetic field dependence of the superconducting transition in the HEA at 103.7 GPa and 179.2 GPa, respectively.

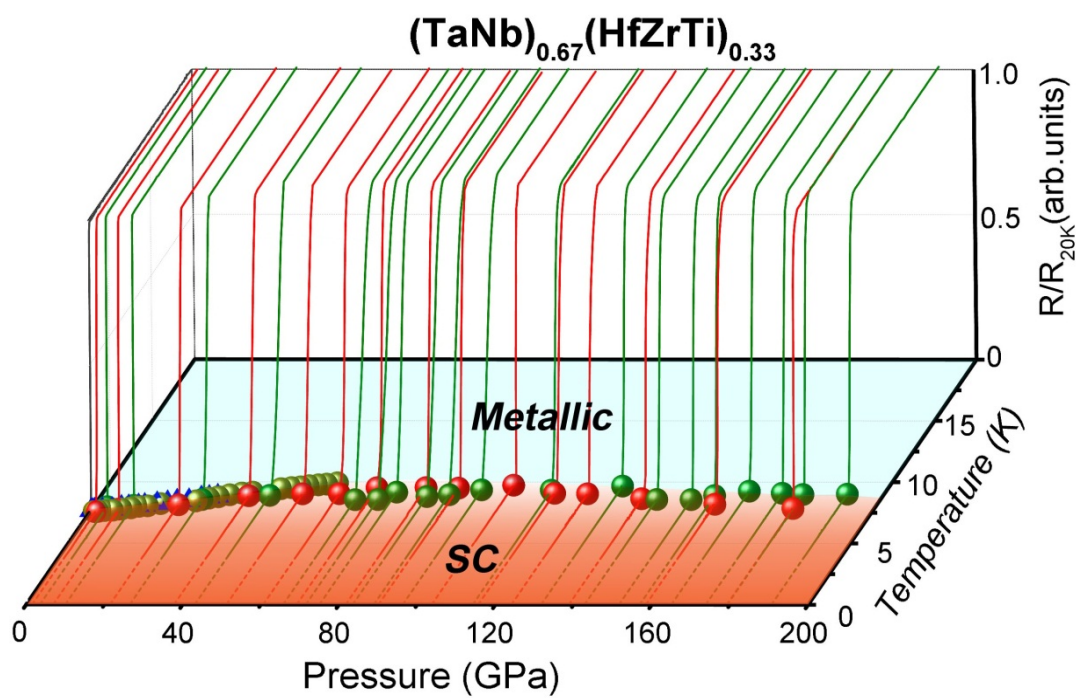

**Figure 3 Phase diagram of superconducting transition temperature vs. applied pressure up to 190.6 GPa for the HEA, combined with plots of the corresponding resistance vs. temperature.** Blue triangles and dark yellow balls in the diagram

represent the zero-resistance superconducting transition temperature $T_C$ obtained from measurements at pressure below 60 GPa, while the red and olive balls stand for zero-resistance $T_C$ from the measurements at pressures up to 179.2 and 190.6 GPa.

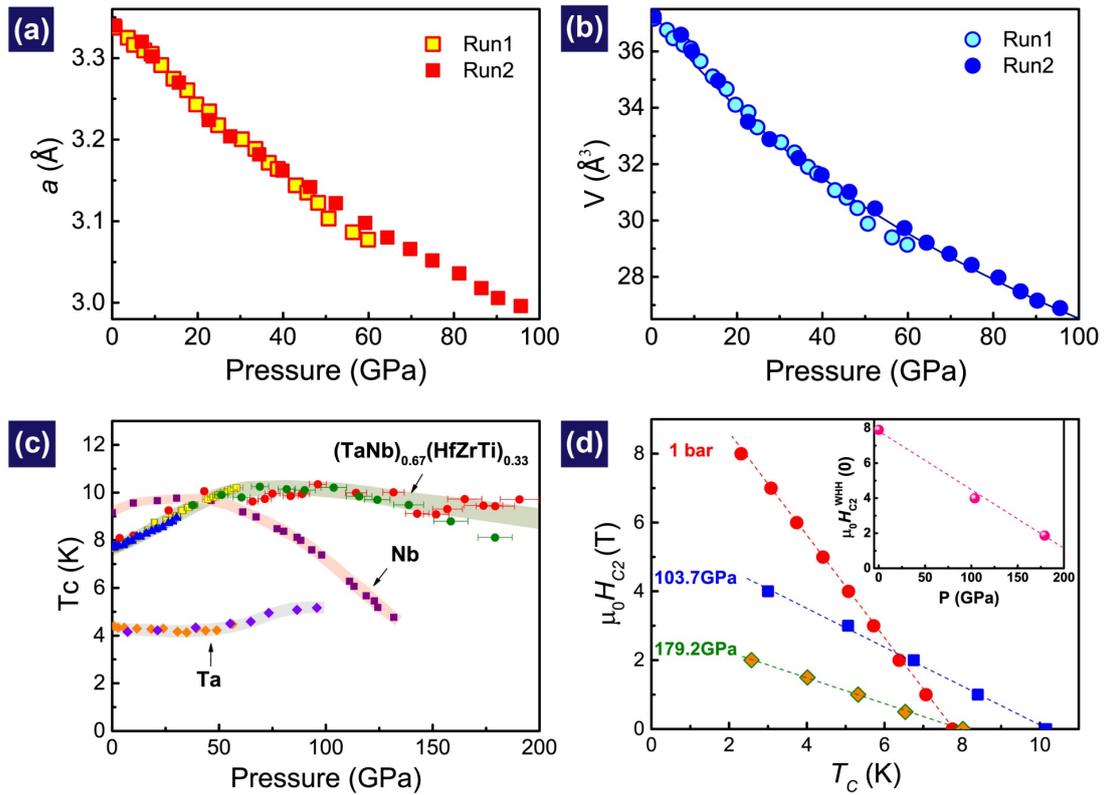

**Figure 4 Details of the high pressure structure and superconductivity information for the superconducting HEA(TaNb)$_{0.67}$(HfZrTi)$_{0.33}$ high entropy alloy and comparison to the behavior of constituent elements.** (a) and (b) Pressure dependence of the lattice parameter and unit cell volume extracted from two independent X-ray diffraction experiments. The standard deviation for the lattice parameters obtained from the diffraction data is approximately 1%. (c) The pressure-dependent change in the superconducting transition temperature of the high

entropy alloy (HEA) compared to those of Nb and Ta, its major elemental constituents. In order to make a better comparison with the reported $T_C$ of elemental Nb (determined by magnetic susceptibility measurements, Ref.28), we use the midpoint $T_C$ of our HEA and Ta samples in panel c. The solid circles, blue triangles and yellow squares are the data for the HEA obtained in this study, the purple squares and orange diamonds are the data previously reported for elemental Nb [Ref.28] and Ta [Ref.29], respectively, and the violet diamonds are the data for elemental Ta found in this study. (d) Superconducting upper critical field $H_{C2}$ as a function of temperature for the HEA at ambient pressure and pressures of 103.7 GPa and 179.2 GPa. The dotted lines represent the Werthamer-Helfand-Hohenberg (WHH) fits. Inset, the pressure dependence of the zero temperature upper critical field $H_{C2}(0)$ for the high entropy alloy superconductor up to ~180 GPa.